\documentclass[preprintnumbers,amsmath,11pt,amssymb,floatfix,superscriptaddress,showpacs,nofootinbib]{revtex4}
\usepackage{graphicx}
\usepackage{epsfig}
\usepackage{bm}
\usepackage{amsfonts}

\input epsf

\begin{document}

\title{\Large New holographic reconstruction of scalar field dark energy models in the framework of chameleon Brans-Dicke cosmology}

\author{\textbf{Surajit Chattopadhyay}}\email{surajcha@iucaa.ernet.in, surajitchatto@outlook.com} \affiliation{Pailan College of Management and Technology, Bengal Pailan Park, Kolkata-700 104, India.}

\author{\textbf{Antonio Pasqua}}\email{toto.pasqua@gmail.com} \affiliation{Department of Physics, University of Trieste, Via Valerio, 2 34127 Trieste, Italy}

\author{\textbf{Martiros Khurshudyan}}\email{khurshudyan@yandex.ru} \affiliation{Department of Theoretical Physics, Yerevan State University, 1 A. Manookian, Armenia.}

\date{\today}

\begin{abstract}
Motivated by the work of Yang et al., \emph{Mod. Phys. Lett. A}, \textbf{26}, 191 (2011), we report a study on the New Holographic Dark Energy (NHDE) model with energy density given by $\rho_D=\frac{3\phi^2}{4\omega}(\mu H^2+\nu \dot{H})$ in the framework of chameleon Brans-Dicke cosmology. We have studied a correspondence between the quintessence, the DBI-essence and the tachyon scalar field models with the NHDE model in the framework of chameleon Brans-Dicke cosmology. Deriving an expression of the Hubble parameter $H$ and, accordingly, $\rho_D$ in the context of chameleon Brans-Dicke chameleon cosmology, we have reconstructed the potentials and dynamics for these scalar field models. Furthermore, we have examined the stability for the obtained solutions of the crossing of the phantom divide under a quantum correction of massless conformally-invariant fields and we have seen that quantum correction could be small when the phantom crossing occurs and the obtained solutions of the phantom crossing could be stable under the quantum correction. It has also been noted that the potential increases as the matter-chameleon coupling gets stronger with the evolution of the universe.
\end{abstract}


\maketitle

\section{Introduction}
 Approaches to account for the late time cosmic acceleration, which is suggested by the two independent observational signals on distant Type Ia Supernovae (SNeIa)  \cite{riess,perlmu,knop}, the Cosmic Microwave Background (CMB) temperature anisotropies measured by the WMAP and Planck satellites \cite{cmb1,cmb2,planck} and Baryon Acoustic Oscillations (BAO) \cite{wmap1,wmap2} fall into two representative categories: in the first,  the concept of ``dark energy" is introduced  in the right-hand side of the Einstein equation in the framework of general relativity (for good reviews see  \cite{copeland-2006,bambareview,DErev}) while in the second one the left-hand side of the Einstein equation is modified, leading to a modified gravitational theory (which is well reviewed in \cite{nojirireview,modreview,modreview1,modreview2}). In a recent review, Bamba et al. \cite{bambareview} demonstrated that both dark energy models and modified gravity theories seem to be in agreement with data and hence, unless higher precision probes of the expansion rate and the growth of structure will be available, these two rival approaches could not be discriminated. Physical origin of dark energy (DE) is one of the largest mysteries not only in cosmology but also in fundamental physics \cite{star1,satro1,star2,star3,copeland-2006}. The cosmological constant $\Lambda$ represents the earliest and the simplest theoretical candidate proposed in order to explain the observational evidence of accelerated expansion. Some tentative deviations from $\Lambda$CDM model may eventually rule out an exact cosmological constant \cite{star4,star5}. A considerable number of models for DE have been proposed till date to explain the late-time cosmic acceleration without the cosmological constant. Such models include a canonical scalar field, the so-called quintessence, a non-canonical scalar field such as phantom, tachyon scalar field motivated by string theories and a fluid with a special equation of state (EoS) called as Chaplygin gas. Other well studied candidates for DE are the k-essence, the quintom and the Agegraphic Dark Energy (ADE) models. Studies on the models previously mentioned include  \cite{copeland-2006,bambareview,DE10,DE11,DE12,DE13,DE14,DE15,DE16,DE17}. There also exists a proposal known as Holographic Dark Energy (HDE) proposed by Li \cite{3}, following the idea that the short distance cut-off is related to the infrared cut-off and it was assumed in \cite{3} that the infrared cut-off relevant to the dark energy is the size of the event horizon. Some notable works on HDE include \cite{holo1,holo2,holo3,holo4}. Furthermore, there exist plethora of literatures on HDE in theoretical aspects as well as observational constraints e.g. \cite{holo5,holo6,holo7}.

The Equation of State (EoS) parameter, defined as $w_{DE}=p_{DE}/\rho_{DE}$  (where $p_{DE}$ and $\rho_{DE}$ denote the pressure and density of DE, respectively), is one of the most important quantity used to describe the features of DE models.  If we restrict ourselves in four-dimensional Einstein's gravity, almost all DE models can be classified according to the behavior of EoS parameter as follow \cite{saridakis}: (i) Cosmological constant: $w_{\Lambda}=-1$; (ii) Quintessence: $w_Q\geqslant -1$; (iii) Phantom: $w_P\leqslant -1$ and (iv) Quintom: its EoS is able to evolve across the cosmological constant boundary. Scalar field models of dark energy are among the most promising and best elaborated ones to match observations of the accelerated expansion of the Universe. The phantom-like behavior of $w_{DE}$ may appear from Brans-Dicke (BD) scalar tensor gravity, from non-standard (negative) potentials, from the non-minimal coupling of scalar Lagrangian with gravity or even usual matter may appear in phantomlike form \cite{nojiriphantom}. Studies devoted to phantom cosmology include \cite{phantom1,phantom2,phantom3,phantom4,phantom5,phantom6} and to quintessence include \cite{quint1,quint2,quint3}.

 In this contribution, we are working on new holographic reconstruction of scalar field dark energy models in the framework of chameleon Brans-Dicke cosmology. In the context of cosmological reconstruction problem, some notable contributions are \cite{recons1,recons2,recons3}. The current work is primarily motivated by \cite{motiv1} and also got inspiration from \cite{motiv2,motiv3}. It is already stated that HDE model is based on the holographic principle, according to which, the number of degrees of freedom of a physical system scales with the area of its boundary \cite{holo1}. Although HDE gives the observational value of DE in the universe and can drive the universe to an accelerated expansion phase, an obvious drawback concerning causality appears in this proposal \cite{motiv1}. In view of this limitation Granda and Oliveros \cite{granda1} proposed a new infrared cut-off for HDE, which is proportional to the square of the Hubble parameter squared $H^2$ and to the time derivative of the Hubble parameter $\dot{H}$, and dubbed this model as new HDE (NHDE) model. The energy density of the NHDE model is given by \cite{granda1}:
 \begin{equation}\label{nhde}
\rho_D=3M_{p}^2 (\alpha H^2+\beta \dot{H}),
 \end{equation}
 where $\alpha$ and $\beta$ are two positive constants. This model could avoid the problem of causality and could solve the coincidence problem \cite{granda1}. In a more recent work, Li et al. \cite{lili} confirmed through action principle that the NHDE model overcomes the causality and circular problems in the original HDE model and putting constraints on the model from the Union$2.1$+BAO+CMB+H$_0$ data \cite{lili} got the goodness of-fit $\chi^2_{min}=548.798$, which they found comparable with the results of the original HDE model $(549.461)$ and the
concordant $\Lambda$CDM model $(550.354)$ and this lead them to conclude that NHDE fit well to the data. Viewing scalar field dark energy models as an effective description of the underlying theory of dark energy, and considering the holographic vacuum energy scenario as pointing in the same direction, Granda and Oliveros \cite{granda2} demonstrated how the scalar field models can be used to describe the holographic energy density as effective theories, for this purpose they studied correspondence between the quintessence, tachyon, K-essence and dilaton energy densities with this NHDE in the flat FRW universe. Connecting these scalar field models with NHDE, they found the explicit forms of the  scalar fields and of the potentials in this reconstruction approach. Karami and Fehri \cite{motiv2} extended the work of \cite{granda2} to non-flat FRW universe i.e. they studied a correspondence between NHDE and quintessence, tachyon, K-essence and dilaton scalar field models in the presence of a spatial curvature and reconstructed scalar field and potential. Sharif and Jawad \cite{sharif2} established a correspondence between the NHDE model and the quintessence, the tachyon, the K-essence and the dilaton scalar field models. In a recent work, Jawad et al. \cite{jawad1} explored holographic reconstruction of modified $f(R)$ Horava-Lifshitz gravity via power-law scale factor and discussed EoS parameter as well as stability of the reconstructed model and remarked about quintessence era in near future with instability.  Dependency of the evolution of equation of state, deceleration parameter and cosmological evolution of Hubble parameter on the parameters of NHDE model were studied in Malekjiani et al \cite{NHDE3}. Considering interaction between dark matter and NHDE, Debnath and Chattopadhyay \cite{NHDE4} investigated the statefinder and the $Om$ diagnostics and also checked the validity of the GSL of thermodynamics with apparent horizon as the enveloping horizon of the universe.

Before demonstrating the current contribution in view of the stated works, let us have a brief overview of the theories of modified gravity as the current contribution is going to explore a cosmological reconstruction in the framework of a modified gravity theory. Modified gravity has become an essential part of theoretical cosmology nowadays \cite{bambareview,nojo,nojo2}. It is proposed as generalization of General Relativity with the purpose to understand the qualitative change of gravitational interaction in the very early and/or very late universe. In particular, it is accepted nowadays that modified gravity may not only describe the early-time inflation and late-time acceleration but also may propose the unified consistent description of the universe evolution epochs sequence: inflation, radiation/matter dominance and dark energy \cite{nojirirecent}. Nojiri and Odintsov \cite{nojo2} summarized the usefulness of modified gravity as follow:
\begin{enumerate}
  \item it provides natural gravitational alternative for dark energy,
  \item it presents very natural unification of the early-time inflation and late-time acceleration thanks
to different role of gravitational terms relevant at small and at large curvature,
  \item it may serve as the basis for unified explanation of dark energy and dark matter.
  \end{enumerate}
Modified gravity models include $f(R)$ gravity (where $R$ is the Ricci scalar curvature) \cite{fr1,fr2,fr3}, $f(T)$ gravity (where $T$ represents the torsion scalar) \cite{ft2,bamba4}, scalar-tensor theories \cite{scalar,scalar1}, braneworld models \cite{brane}, Galileon gravity \cite{gali}, Gauss-Bonnet gravity \cite{gb} and so on. Bamba et al. \cite{bamba3} investigated the future evolution of the dark energy universe in modified gravities including $f(R)$ gravity, string-inspired scalar-Gauss-Bonnet and modified Gauss-Bonnet ones and an ideal fluid with inhomogeneous equation of state and constructed several examples of the modified gravity that produces accelerating cosmologies ending at the finite-time future singularity by applying the reconstruction program. Cosmological evolution of the equation of state for dark energy $w_{DE}$ in the exponential and logarithmic as well as their combination in the framework of $f(T)$ theories was studied in Bamba et al. \cite{bamba4}. A reconstruction scheme for modified gravity realizing a crossing of the phantom divide was proposed in Bamba et al. \cite{bamba5}. Appearance of finite-time future singularities in $f(T)$ gravity was demonstrated in Bamba et el. \cite{bamba6}. In another work, Bamba et al \cite{bamba7} explored the cosmological evolution in a modified gravity $f(R)=R +c_1 (1-e^{- c_2 R})$ and demonstrated that the late-time cosmic acceleration following the matter-dominated stage can be realized n that model. Bamba \cite{bamba8} showed that the crossing of the phantom divide can be realized in the combined $f(T)$ theory constructed with the exponential and logarithmic terms.

Recently, various scalar tensor theories have been considered extensively and one important example of the scalar tensor theories is the Brans-Dicke (BD) theory of gravity which was introduced by Brans and Dicke \cite{BD0} to incorporate the Mach principle in the Einstein's theory of gravity.  BD theory is proposed as the natural extension of Kaluza-KLein idea of unification \cite{momeni1}. BD parameter has some interesting properties as a candidate of DE when it has been studied in the non minimally coupled regime \cite{momeni1,momeni2,momeni3,momeni4,momeni4-1,momeni5}. The popularity of BD modified gravity \cite{BD1,BD2} lies in the fact that it naturally arises as the low energy limit of many other quantum gravity theories, like the Kaluza-Klein one or the superstring theory. In this paper, we decided to consider the New Holographic Dark Energy (NHDE) model in the framework of the chameleon Brans-Dicke modified gravity theory, in which there is a non-minimal coupling between the matter field and the scalar field $\phi$ which is usually known in literature with the name of chameleon field \cite{19,18} since its main  physical properties strongly depend  on the environment. Waterhouse \cite{34} derived that the deviations from Newtonian gravity due to the chameleon field of the Earth are suppressed by nine orders of magnitude by the thin-shell effect. In a recent work, Chattopadhyay \cite{surajit} studied the matter-chameleon coupling considering extended holographic Ricci Dark Energy model in chameleon Brans-Dicke cosmology. Instead, Bisbar \cite{bisbar} considered a generalized Brans-Dicke model in which the scalar field has a potential function and it can couple non-minimally with the matter sector. Late-time dynamics of a chameleonic generalized Brans-Dicke cosmology with a power law chameleonic function has been well studied in a recent paper of El-Nabulsi \cite{Rami}. Other important works on chameleon gravity have been done in \cite{32,7,1,2,23mota,mionade}.

 The present contribution is organized as follow: in Section II, we study the main cosmological properties of the NHDE model in the framework of Brans-Dicke chameleon cosmology. In Section III, we make a correspondence between the reconstructed NHDE model and three differen scalar field models, i.e. the quintessence, the DBI-essence and the tahcyon scalar field models. Finally, in Section IV, we write the conclusions of this paper.

\section{NHDE  MODEL in chameleon BD cosmology}
We begin this Section with the description of the main cosmological properties of the chameleon Brans-Dicke (BD) theory.\\
According to BD theory, the scalar field is coupled non-minimally to the matter field via the action $S$ given by \cite{bisbar}:
\begin{equation}\label{action}
S=\frac{1}{2}\int d^4x\sqrt{-g}\left(\phi
R-\frac{\omega}{\phi}g^{\mu\nu}\nabla_{\mu}\phi\nabla_{\nu}\phi-2V+2f(\phi)L_m\right)
\end{equation}
where $R$ indicates the Ricci scalar curvature, $\phi$ represents the Brans-Dicke scalar field with an associate potential $V \left(\phi\right)$, $\omega$ indicates the dimensionless Brans-Dicke parameter, $g^{\mu \nu}$ represents the metric tensor with determinant given by $g$, $L_m$  represents the matter Lagrangian and, finally, $f\left(  \phi \right)$ represents an arbitrary function of the scalar field $\phi$. The last term in the action $S$ given in Eq. (\ref{action}) represents the term which gives us information about the interaction between the matter Lagrangian and the arbitrary function $f\left( \phi \right)$. We must also emphasize here that, in the limiting case corresponding to $f\left(  \phi \right)= 1$, we obtain the standard BD cosmology.\\
Varying the action $S$ given in Eq. (\ref{action}) with respect to the metric tensor $g_{\mu\nu}$ and $\phi$, we obtain the following field equations:
\begin{eqnarray}\label{field1}
\phi G_{\mu\nu}&=&T_{\mu\nu}^\phi+f(\phi)T_{\mu\nu}^m,\\
(2\omega+3)\Box \phi+2(2V-V'\phi)&=&T^m f-2f'\phi_m, \label{field2}
\end{eqnarray}
where $G_{\mu\nu}$ is the Einstein tensor, $\Box=\nabla^{\mu}\nabla_\mu$ (with $\nabla_\mu$
representing the covariant derivative) $T^m=g^{\mu\nu}T_{\mu\nu}^m$
and the prime denotes a differentiation with respect to $\phi$. In Eq. (\ref{field1}), we have that:
\begin{equation}\label{field11}
T_{\mu\nu}^\phi=\frac{\omega}{\phi}\left(\nabla_{\mu}\phi
\nabla_{\nu}\phi-\frac{1}{2}g_{\mu\nu}\nabla_{\alpha}\phi\nabla^{\alpha}\phi\right)+\left(\nabla_{\mu}\nabla_{\nu}\phi-g_{\mu\nu}\Box
 \phi \right)-V(\phi)g_{\mu\nu},
\end{equation}
and
\begin{equation}\label{field12}
T_{\mu\nu}^m=\frac{-2}{\sqrt{-g}}\frac{\delta(\sqrt{-g}L_m)}{\delta
g^{\mu\nu}}.
\end{equation}
Because of the explicit coupling between matter system and $\phi$,
the stress tensor $T_{\mu\nu}^m$ is not divergence free.
We now apply the above framework to a homogeneous and
isotropic universe described by the Friedman-Robertson-Walker
metric given by:
\begin{equation}\label{FRW}
ds^2=-dt^2+a^2(t)\left(\frac{dr^2}{1-kr^2}+r^2 d\Omega^2\right).
\end{equation}
The universe is open, closed or flat according as
$k=-1,~+1~\textrm{or}~0 $. Moreover, we have that $a\left(t\right)$ represents the scale factor (which gives information about the expansion of the universe), $r$ gives the radial component of the metric, $t$ indicates the cosmic time and $d\Omega^2 = r^2 \left(d\theta ^2 + \sin^2 \theta d\varphi ^2\right) $ denotes the solid angle element (squared).
$\theta$ and $\varphi$ are the usual azimuthal and polar angles, with $0\leq \theta \leq \pi$ and $0\leq \varphi \leq 2\pi$.
The coordinates $\left(r , t, \theta, \varphi  \right)$ are known as comoving coordinates.\\
In a spatially flat universe (i.e. for $k=0)$, Eqs.
(\ref{field1}) and (\ref{field12}) yield \cite{bisbar}:
\begin{eqnarray}\label{fried1}
3H^2&=&\frac{f}{\phi}\rho+\frac{\omega}{2}\frac{\dot{\phi}^2}{\phi^2}-3H
\frac{\dot{\phi}}{\phi}+\frac{V}{\phi}, \\
\label{fried2}
3(\dot{H}+H^2)&=&-\frac{3\rho}{\phi(2\omega+3)}\left\{\gamma \phi
f'+\left[\omega\left(\gamma+\frac{1}{3}\right)+1\right]f\right\} \nonumber \\
&&-\omega\frac{\dot{\phi}^2}{\phi^2}+3H
\frac{\dot{\phi}}{\phi}+\frac{1}{2\omega+3}\left[3V'+(2\omega-3)\frac{V}{\phi}\right], \\
\label{fried3}
(2\omega+3)(\ddot{\phi}+3H\dot{\phi})&-&2(2V-\phi
V')=\rho\left[(1-3\gamma)f+2\gamma\phi f'\right].
\end{eqnarray}
We must underline here that a dot indicates a time derivative while the prime indicates a derivative with respect to $\phi$.\\
In this paper, our purpose is to generalize the work of Karami and Fehri \cite{motiv2}
to the NHDE model with energy density $\rho_D$ given by \cite{motiv1}:
\begin{eqnarray}
\rho_D=\frac{3 \phi^2 }{4 \omega }\left(\mu H^2  + \nu \dot{H} \right), \label{NHDE}
\end{eqnarray}
where $\mu$ and $\nu$ are two constant parameters and the overdot represents the first time derivative.

We consider the following ansatz for $\phi$, $V$ and $f$ \cite{bisbar,motiv1}:
\begin{eqnarray}
\phi&=&\phi_0 a^\alpha, \nonumber\\
V&=&V_0 \phi^\beta, \nonumber\\
f&=&f_0 \phi^\gamma, \nonumber \label{ansatz}
\end{eqnarray}
where $\alpha,~\beta$ and $\gamma$ are constant
parameters and $\phi_0,~V_0$ and $f_0$ are positive quantities representing the present day values of the corresponding quantities. One attribute of taking this kind of ansatz is scale invariance of power laws. Given a relation in power law form, scaling the argument by a constant factor  $c$ causes only a proportionate scaling of the function itself. Issues related to power-law choice of potential are discussed in \cite{power1,power2}.\\
Using the aforesaid ansatz in Eq. (\ref{fried1}) we get the following differential equation for $H^2$:
\begin{equation}\label{diffeqn}
\frac{dH^2}{da}+\left[\frac{2
\mu }{\nu }-\frac{8 e^{\alpha x} \phi_0 \left(e^{\alpha x } \phi_0\right)^{-2-\nu } \omega }{3 f_0 \nu
}\right] H^2=-\frac{8 \left(e^{\alpha x} \phi_0\right)^{-2+\beta -\nu }V_0 \omega }{3 f_0 \nu }.
\end{equation}
Solving Eq. (\ref{diffeqn}), we get the expression of reconstructed $\tilde{H}^2$ as a function of the scale factor $a$ as follow:
\begin{equation}
\begin{array}{c}
\tilde{H}^2(a)=\left(\frac{3}{8}\right)^{\eta _4} a^{\alpha  (-2+\beta -\nu )} e^{-a^s \eta _6} \eta _1 \eta _3 \left(-a^s
\eta _5\right)^{\eta_4} \Gamma\left[\eta _2,a^s \eta _5\right], \label{Hsqr}
\end{array}
\end{equation}
where:
\begin{eqnarray}
 \eta_1&=&\frac{\left(\frac{3}{8}\right)^{-1+\frac{2 \mu }{s \nu }-\frac{\alpha  (2-\beta +\nu )}{s}}}{f_0 s \nu },\\
 \eta_2&=&\frac{2 \mu +\alpha  (-2+\beta -\nu ) \nu }{s \nu },\\
 \eta_3&=&\phi_0^{-2+\beta -\nu } V_0 \omega ,\\
\eta_4&=&-\frac{2 \mu }{s \nu }+\frac{\alpha  (2-\beta +\nu )}{s},\\
\eta_5&=&-\frac{8 \phi_0^{-1-\nu } \omega }{3 f_0 \alpha  \nu  (1+\nu )},\\
\eta_6&=&-\frac{8 \left(a^{\alpha } \phi_0\right)^{-\nu } \omega }{3 f_0 \phi_0 \alpha  \nu  (1+\nu )},\\
s&=&-\alpha \left(1+\nu\right).
\end{eqnarray}
Moreover, we have that in Eq. (\ref{Hsqr}) $\Gamma$ represents the Gamma function.\\
Subsequently, using the relation $\dot{H}=\frac{a}{2}\frac{dH^2}{da}$, we obtain the following relation for $\dot{\tilde{H}}(a)$:
\begin{equation}
\begin{array}{c}
\dot{\tilde{H}}(a)=-2^{-1-3 \eta_4} 3^{\eta_4} a^{\alpha  (-2+\beta -\nu )} e^{-a^s (\eta_5+\eta_6)} \eta_1
\eta_3 \left(-a^s \eta_5\right)^{\eta_4} \left(s \left(a^s \eta_5\right)^{\eta_2}-e^{a^s \eta_
5} \left(s \left(\eta_4-a^s \eta_6\right)\right.\right.\\
\left.\left.+\alpha  (-2+\beta -\nu )\right) \Gamma\left[\eta_2,a^s \eta_5\right]\right) \label{Hdot}
\end{array}
\end{equation}
We can use Eq. (\ref{Hsqr}) and (\ref{Hdot}) in Eq. (\ref{NHDE}) to reconstruct the density of the NHDE in chameleon BD cosmology, obtaining:
\begin{equation}
\begin{array}{c}\label{density}
\rho_D(a)=-\frac{2^{-3-3 \eta_4} 3^{1+\eta_4}}{\omega }
\left[a^{\alpha  (\beta -\nu )} e^{-a^s
(\eta_5+\eta_6)} \phi_0^2 \eta_1 \eta_3 \times\right. \\
\left. \left(-a^s \eta_5\right)^{\eta_4} \left\{s \left(a^s
\eta_5\right)^{\eta_2} \nu -e^{a^s \eta_5} \left(2 \mu -\nu  \left(-s \eta_4+a^s s \eta_6+\alpha  (2-\beta
+\nu )\right)\right) \Gamma\left[\eta_2,a^s \eta_5\right]\right\}\right]
\end{array}
\end{equation}
Since we are assuming that the universe is filled with NHDE, the conservation equation for $\rho_D$ is given by the following relation:
\begin{equation}\label{conserve}
\dot{\rho}_D+3H\rho_D\left(1+w_D\right)=0,
\end{equation}
where $w_D$ represents the EoS parameter of DE.\\
From Eqs. (\ref{density}) and (\ref{conserve}), we can derive the reconstructed equation of state (EoS) parameter $w_D$ as a function of the scale factor $a$ as follow:
\begin{equation}
\begin{array}{c}\label{EoS}
w_D(a)=-1-\{s \left(a^s \eta_5\right)^{\eta_2} \left(-2 \mu +\nu  \left(-s \left(\eta_2+2 \eta_4-a^s (\eta_5+2 \eta_6)\right)\right.\right.\\
\left.\left.+2 \alpha  (1-\beta +\nu )\right)\right)+e^{a^s \eta_5} \left(a^{2 s} s^2 \eta_6^2 \nu +(s \eta_4+\alpha  (\beta -\nu )) (2 \mu +(s \eta_4+\alpha  (-2+\beta -\nu )) \nu )\right.\\
\left.-a^s s \eta_6 (2 \mu +(s+2 s \eta_4+2 \alpha
 (-1+\beta -\nu )) \nu )\right) \Gamma\left[\eta_2,a^s \eta_5\right]\}\times\\
\left\{3 \left(s \left(a^s \eta_5\right)^{\eta_2} \nu -e^{a^s \eta_5} \left(2 \mu -\nu  \left(-s \eta_4+a^s s \eta_6+\alpha  (2-\beta +\nu )\right)\right) \Gamma\left[\eta_2,a^s \eta_5\right]\right)\right\}^{-1}
\end{array}
\end{equation}

\begin{figure}[ht] \begin{minipage}[b]{0.45\linewidth} \centering\includegraphics[width=\textwidth]{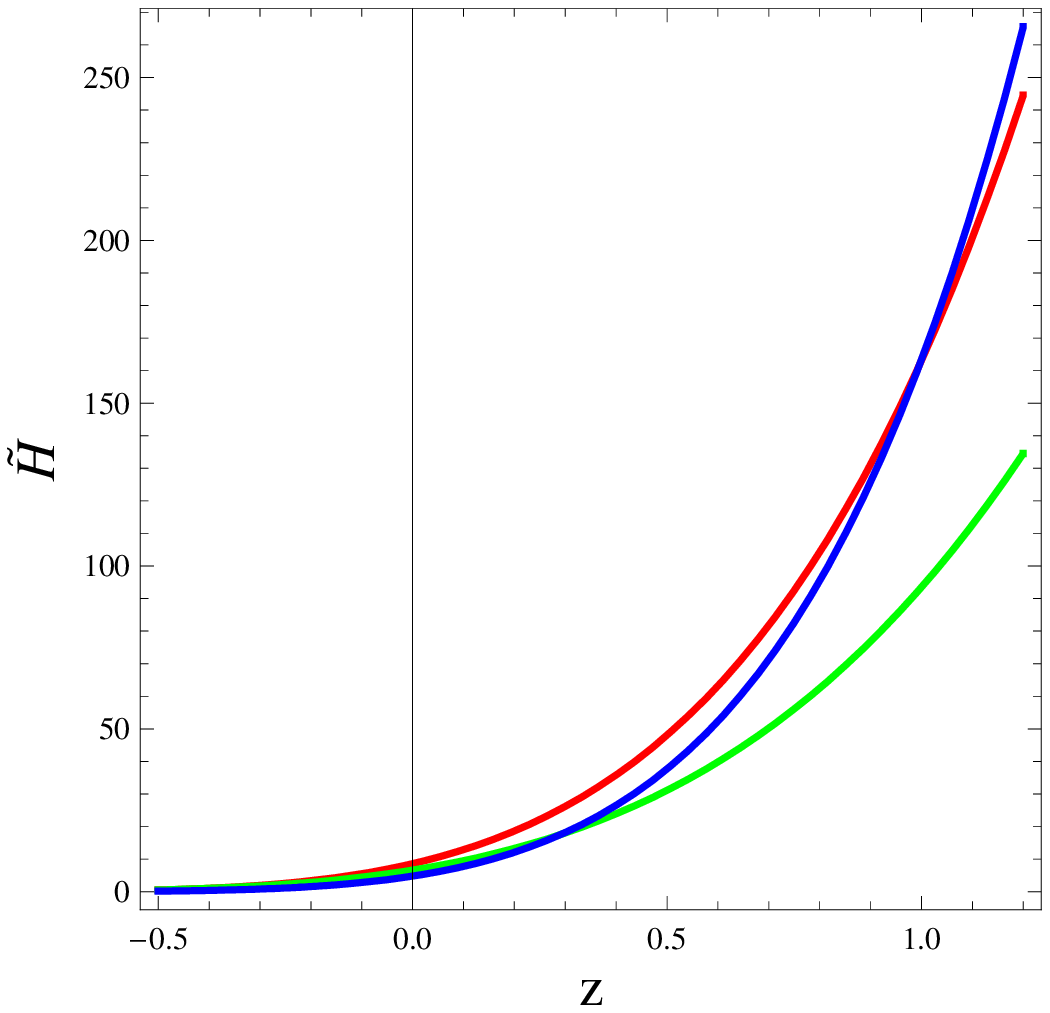} \caption{Plot of the reconstructed Hubble parameter $\tilde{H}$ obtained in Eq. (\ref{Hsqr}).} \label{fig1} \end{minipage} \hspace{0.5cm} \begin{minipage}[b]{0.45\linewidth} \centering\includegraphics[width=\textwidth]{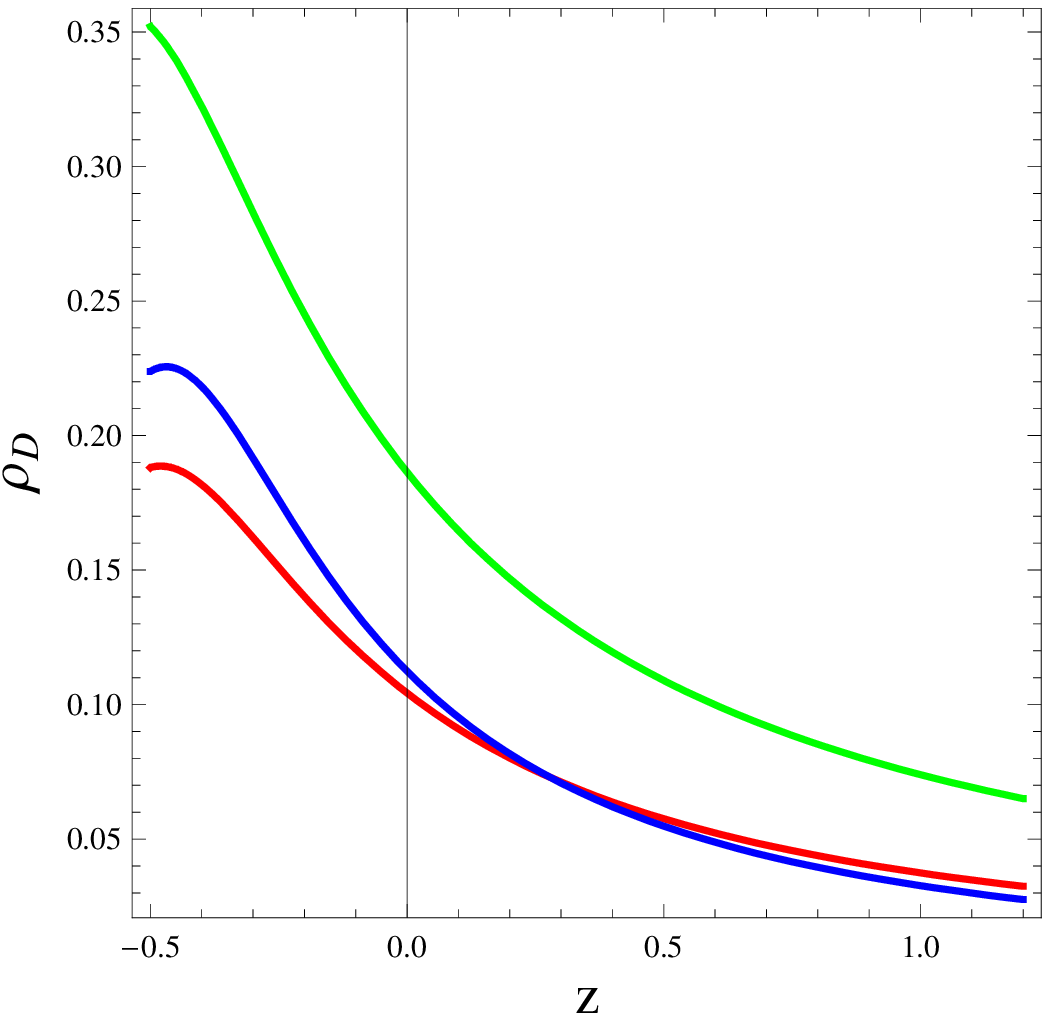} \caption{Plot of the reconstructed energy density $\rho_D$.\\ See Eq. (\ref{density})} \label{fig2} \end{minipage} \hspace{0.5cm} \begin{minipage}[b]{0.45\linewidth} \centering\includegraphics[width=\textwidth]{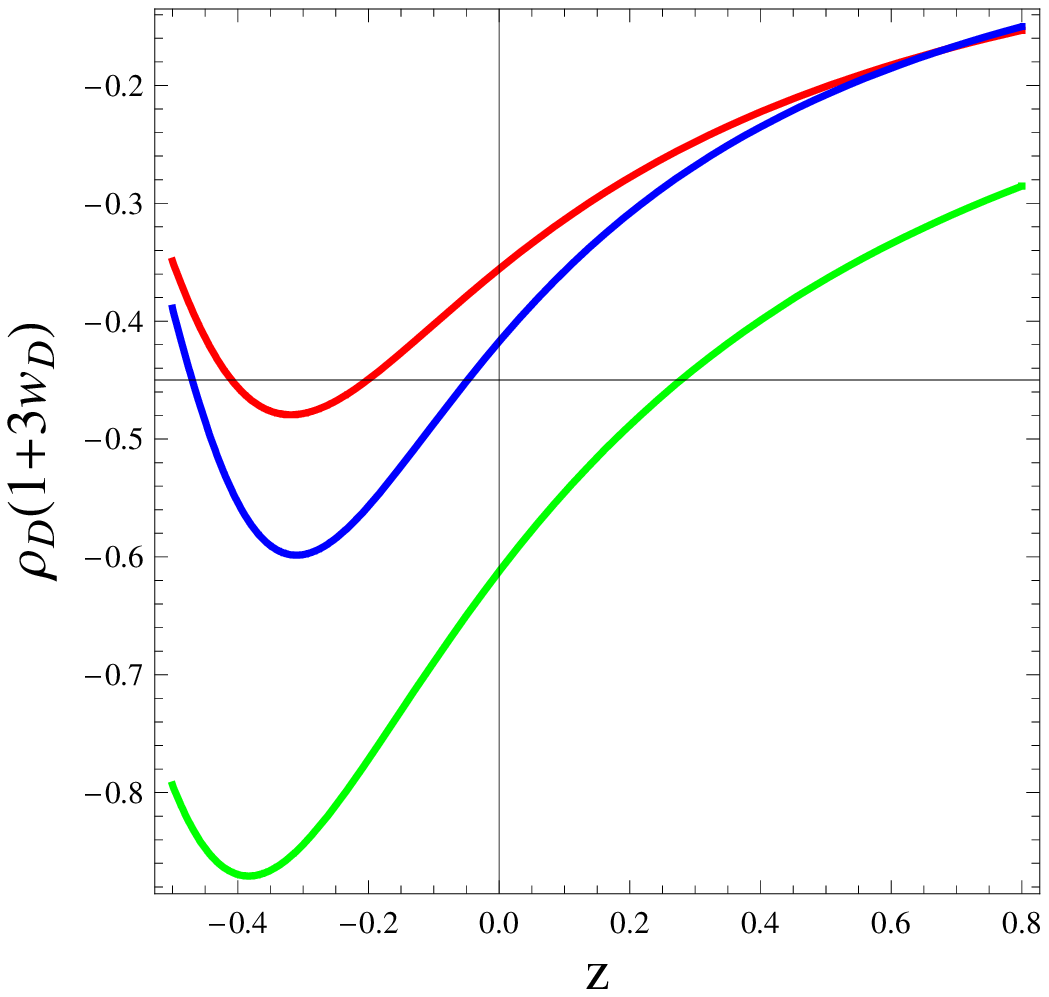}\caption{Plot of the evolution of $\rho_D(1+3w_D)$ (strong energy condition test). See Eqs. (\ref{density}) and (\ref{EoS}).} \label{fig3} \end{minipage}\hspace{0.5cm}\begin{minipage}[b]{0.45\linewidth} \centering\includegraphics[width=\textwidth]{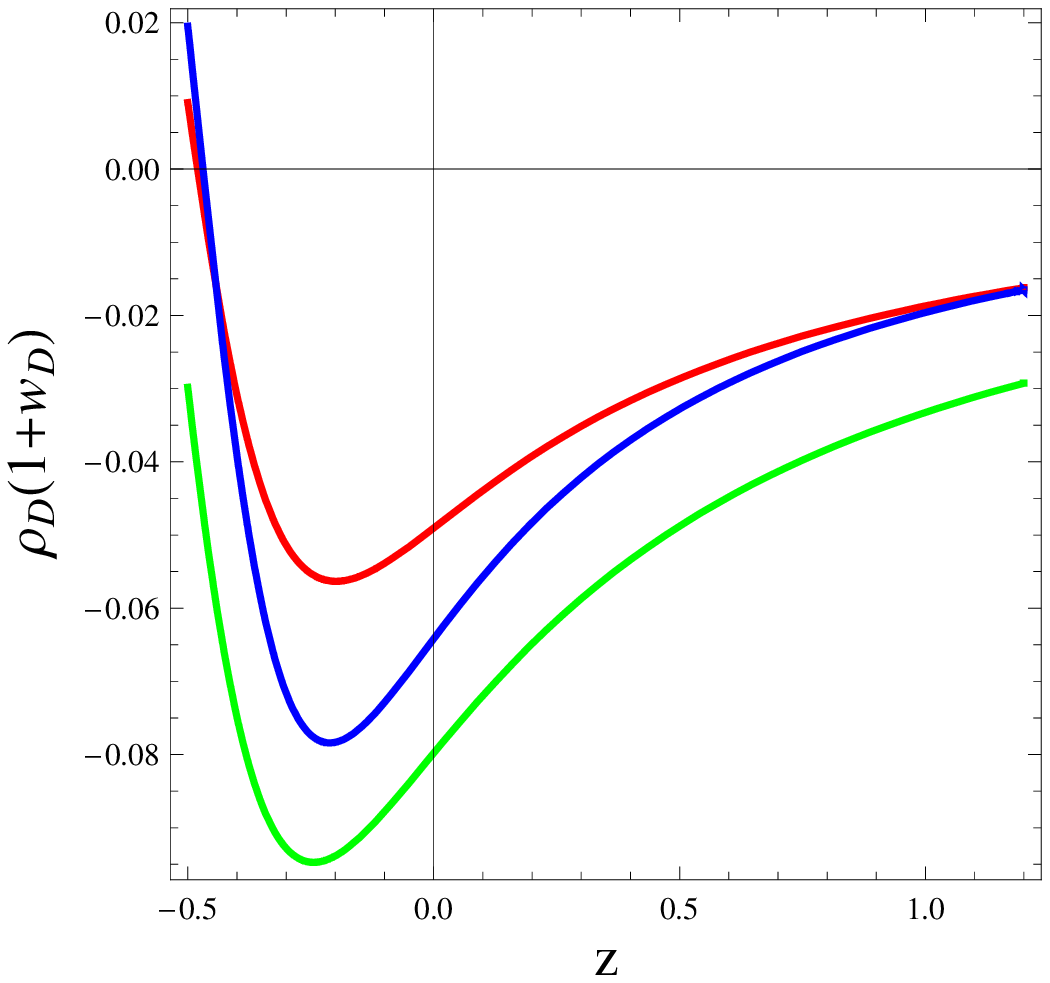}\caption{Plot of the evolution of $\rho_D(1+w_D)$ (null energy condition test). See Eqs. (\ref{density}) and (\ref{EoS}).} \label{fig4}\end{minipage}\end{figure}

 We now plot the reconstructed cosmological parameters against the redshift $z$. In all Figures, we have that red, green and blue lines correspond to $\{\mu=0.65,~\nu=0.20\}$, $\{\mu=0.60,~\nu=0.25\}$ and $\{\mu=0.55,\nu=0.15\}$, respectively. For all other figures, the other parameters present in the equations we derived are set as $\alpha=5,~\beta=-0.7,~\gamma=-0.9,~\phi_0=0.12,~\omega=-\frac{3}{2}+10^{-22},~f_0=1,~V_0=2$. The choice of the value of the BD parameter $\omega$ is based on ref. \cite{Hrycyna}. Observational results coming from SNeIa data suggest a range of possible values for the EoS parameter of $-1.67<w<-0.62$ \cite{MNR}. Using a set of variations in the values of $\mu$ and $\nu$ in Eq. (\ref{EoS}) we find that the results are in good agreement with \cite{MNR}. In Table I, we are showing a set of values of the reconstructed EoS based on the chosen values of the parameters. It is apparent from this table that the EoS parameter $-1.67<w<-0.62$.
\begin{table}
\begin{center}
\caption{Values of reconstructed $w_D$ (Eq. (\ref{EoS})) for different choices of parameter values.}
    \begin{tabular}{ | p{5.2cm} | p{3cm} | p{3cm} | p{3cm} |}
    \hline
    Choice of $\alpha$, $\beta$ and $\gamma$ & vales of $w_D~(\mu=0.55,~\nu=0.15)$ & values of $w_D~(\mu=0.65,~\nu=0.20)$ & values of $w_D~(\mu=0.60,~\nu=0.25)$ \\ \hline\hline
    $(\alpha=5,~\beta=-0.7,~\gamma=-0.9)$ & -1.47265 & -1.47040 & -1.47797 \\ \hline
    $(\alpha=4.5,~\beta=-0.6,~\gamma=-1)$ & -1.57145 & -1.56864 & -1.57484 \\ \hline
   $(\alpha=4.1,~\beta=-0.8,~\gamma=-1.1)$ & -1.25035& -1.24863 & -1.25514 \\
    \hline
    \end{tabular}
\end{center}
\end{table}
In Fig. \ref{fig1}, we observe the evolution of the reconstructed Hubble parameter $\tilde H$ with redshift $z=a^{-1}-1$. We observe a decaying pattern of $\tilde H$ with evolution of the universe. This in consistent with the accelerated expansion of the universe. In Fig. \ref{fig2}, the reconstructed NHDE density is plotted and indicates its dominance with the evolution of the universe. This is consistent with the current dark-energy dominated era.\\
We now want to have a deeper look into the Eq. (\ref{EoS}) derived through the reconstructed $\tilde H$. In Eq. (\ref{EoS}), the expression $3 \left(s \left(a^s \eta_5\right)^{\eta_2} \nu -e^{a^s \eta_5} \left(2 \mu -\nu  \left(-s \eta_4+a^s s \eta_6+\alpha  (2-\beta +\nu )\right)\right) \Gamma\left[\eta_2,a^s \eta_5\right]\right)~\neq0$. Considering the forms of $\eta_X~(X=1,2,..,6)$ it can be verified that the above expression is surely  positive if $\alpha<0,~~-1<\nu<0$ or $\alpha>0,~~\nu<-1$. Since we have taken $\phi=\phi_0 a^{\alpha}$, we have $\dot{\phi}=\alpha \phi H$. If $\alpha<0$, then the scalar field decays with the evolution of the universe.

Since the behavior of the remaining part of the expression is too complicated to make any inference on its ``quintessence" or ``phantom"-like behavior, we only depend on the \emph{strong} and \emph{null} energy conditions to put some light into the behavior of the equation of state parameter. Null energy condition is satisfied if $\rho_D(1+w_D)\geq 0$ while the strong energy condition condition is satisfied if $\rho_D(1+w_D)\geq 0~~\text{and}~~\rho_D(1+3w_D)\geq 0$. Figs. \ref{fig3} and \ref{fig4} indicate that both of the energy conditions are violated. This indicates ``phantom"-like behavior of the equation of state parameter.

In this Section, we reconstructed the NHDE in the framework of chameleon BD cosmology. In the following Section, our objective is to study the correspondence between this reconstructed NHDE model and the scalar field models, namely (i) Quintessence dark energy, (ii) DBI-essence dark energy and (iii) Tachyon dark energy.

\subsection{Stability under a quantum correction}
Following \cite{bamba5,bambaquantum} we examine the stability for the obtained solutions of the crossing of
the phantom divide under a quantum correction of massless conformally-invariant fields. Quantum effects produce the conformal anomaly \cite{bamba5}:
\begin{equation}\label{quantum1}
  T_A=b\left(F+\frac{2}{3}\square R\right)+b' G+b''\square R,
\end{equation}
where
\begin{eqnarray}
F&=&\frac{1}{3}R^2-2R_{ij}R^{ij}+R_{ijkl}R^{ijkl},\\
G&=&R^2-4R_{ij}R^{ij}+R_{ijkl}R^{ijkl}.
\end{eqnarray}
In the FRW universe, we have that:
\begin{eqnarray}
F&=&0,\\
G&=&24(\dot{H}H^2+H^4).
\end{eqnarray}
For $N$ real scalar, $N_{1/2}$ Dirac spinor, $N_1$ vector fields, $N_2(=0 ~\textrm{or}~1)$ gravitons and $N_{HD}$ higher derivative conformal scalars, we have the following expressions for $b$ and $b'$:
\begin{eqnarray}
b&=&\frac{N+6N_{1/2}+12N_{1}+611 N_2-8N_{HD}}{120(4\pi)^2},~~\\
b'&=&-\frac{N+11N_{1/2}+62N_1+1411N_2-28N_{HD}}{360(4\pi)^2},
\end{eqnarray}
with $b''$ which can be arbitrary. If we assume that $T_A$ can be given by the effective energy density $\rho_A$ and pressure $p_A$ from the conformal anomaly as:
\begin{equation}\label{quantum2}
\dot{\rho}_A+3H(\rho_A+p_A)=0,
\end{equation}
the following expressions for $\rho_A$ and $p_A$ can be found:
\begin{eqnarray}
\rho_A&=&-\frac{1}{a^4}\int dta^4HT_A, \label{TA}\\
p_A&=&-\frac{1}{3a^4}\int dta^4HT_A+\frac{T_A}{3}. \label{TA2}
\end{eqnarray}
At phantom crossing, we must have $\dot{H}=0$. If we assume that the magnitude of the Hubble rate $H$ could be the order of
the present Hubble constant $H_0$, then for phantom crossing we have:
\begin{eqnarray}
\rho_D=\frac{3\phi^2}{4\omega}\tilde{H}_0^2\mu
\end{eqnarray}
where $\tilde{H}_0^2=\tilde{H}^2|_{(a=1)}=\left(\frac{3}{8}\right)^{\eta _4}  e^{- \eta _6} \eta _1 \eta _3 \left(-
\eta _5\right)^{\eta_4} \Gamma\left[\eta _2, \eta _5\right]$.

The results obtained in Eqs. (\ref{TA}) and (\ref{TA2}) tell that we may assume $\rho_A\sim p_A\sim T_A$. Then, we find \cite{bambaquantum}:
\begin{eqnarray}\label{TAA}
\rho_A\sim p_A\sim C\tilde{H}_0^4
\end{eqnarray}
where $C$ represents a dimensionless constant of the order of $\sim 10^{2\sim 3}$.
Thus, for the reconstructed model, we obtain:
\begin{eqnarray}\label{TAA}
\frac{\rho_D}{\rho_A}=\frac{3\phi^2\mu}{4C\omega \tilde{H}_0^2}\approx 10^{19}\frac{\mu}{\tilde{H}_0^2},
\end{eqnarray}
hence we can conclude that
\begin{eqnarray}
|\rho_D|\gg |\rho_A|.
\end{eqnarray}
Therefore, the quantum correction could be small when the phantom crossing occurs and the
obtained solutions of the phantom crossing in this paper could be stable under the quantum
correction.
\section{New holographic reconstruction of scalar field models in BD cosmology}
Sahni and Starobinsky \cite{satro1} discussed various aspects of reconstructing the expansion history of the Universe
and to probe the nature of dark energy. Below, we will study the correspondence between NHDE model and
the quintessence, the DBI-essence and the tachyon scalar field models in the framework of flat chameleon Brans-Dicke universe. We will also reconstruct the potentials and the dynamics for these scalar field models. We can give the related results of scalar fields and potentials for the NHDE model in the flat chameleon Brans-Dicke universe. In order to establish this correspondence, we compare the energy density of the NHDE model given in Eq. (\ref{density}) with the corresponding energy density of the scalar field model, and we also equate the EoS for these scalar models with the EoS for the NHDE model given in Eq. (\ref{EoS}). We must also emphasize here that  we indicate the scalar field with $\varphi$ in order to differ it from the scalar field $\phi$ in Brans-Dicke theory.
\subsection{Reconstruction of quintessence dark energy model}

Quintessence is a dynamical, evolving, spatially inhomogeneous component with
negative pressure. Unlike a cosmological constant, the quintessential
pressure and energy density evolve with the time and the EoS parameter may also do so. A common model of quintessence is the energy density associated with a scalar
field $Q$ slowly rolling down a potential $V(Q)$. A detailed discussion on quintessence dark energy is available in the review \cite{quint0}. The energy density $\rho_Q$ and pressure $p_Q$ of the quintessence scalar field $\varphi$ are given, respectively, by \cite{quint1,quint2,quint3}:
\begin{eqnarray}
\rho_Q&=&\frac{1}{2}\dot{\varphi}^2+V(\varphi),\\
p_Q&=&\frac{1}{2}\dot{\varphi}^2-V(\varphi).
\end{eqnarray}
Moreover, the EoS parameter can be written as follow:
\begin{eqnarray}
w_Q= \frac{p_Q}{\rho_Q} = \frac{\dot{\varphi}^2-2V(\varphi)}{\dot{\varphi}^2+2V(\varphi)}.
\end{eqnarray}

As we are reconstructing the quintessence model based on NHDE in the framework of chameleon BD cosmology, we shall consider $\rho_Q=\rho_D$ and $w_Q=w_D$. Hence, we have:
 \begin{eqnarray}
\dot{\varphi}^2&=&\rho_D(1+w_D),\\
V(\varphi)&=&\frac{\rho_D}{2}(1-w_D),
\end{eqnarray}
where $\rho_D$ and $w_D$ are given in Eqs. (\ref{density}) and (\ref{EoS}), respectively. Based on the reconstructed Hubble parameter, we express $\dot{\varphi}^2$ and $V(\varphi)$ as functions of the scale factor $a$ as follow:

\begin{eqnarray}
\label{qfield}
\dot{\varphi}(a)^2&=&-\frac{1}{\omega }3^{\eta_4} 8^{-1-\eta_4} a^{\alpha  (\beta -\nu )} e^{-a^s (\eta_5+\eta_6)} \phi_0^2
\eta_1 \eta_3 \left(-a^s \eta_5\right)^{\eta_4}
 \left(s \left(a^s \eta_5\right)^{\eta_2} \right. \nonumber\\
 &&\left.\left(-2
\mu +\nu  \left(-s \left(\eta_2+2 \eta_4-a^s (\eta_5+2 \eta_6)\right)+\right.\right.\right.\nonumber \\
&&\left.\left.\left.\alpha  (2-2 \beta +2 \nu )\right)\right)+e^{a^s
\eta_5} \left(\alpha ^2 \nu  \left(\beta ^2-2 \beta  (1+\nu )+\nu  (2+\nu )\right)+s \left(s \eta_4^2 \nu +2 \eta_4 \left(\mu
-a^s s \eta_6 \nu \right)+\right.\right.\right.\nonumber\\
&&\left.\left.\left.a^s \eta_6 \left(-2 \mu +s \left(-1+a^s \eta_6\right) \nu \right)\right)+2 \alpha  \left(\beta
 \left(\mu +s \left(\eta_4-a^s \eta_6\right) \nu \right)-\nu  \left(\mu +s \left(\eta_4-a^s \eta_6\right) (1+\nu
)\right)\right)\right) \right. \nonumber \\
&& \left. \times \Gamma\left[\eta_2,a^s \eta_5\right]\right),
\end{eqnarray}

\begin{eqnarray}
\label{qpotential}
V(\varphi(a))&=&-\frac{2^{-4-3 \eta_4} 3^{1+\eta_4}}{\omega } a^{\alpha  (\beta -\nu )} e^{-a^s (\eta_5+\eta_6)}
\phi_0^2 \eta_1 \eta_3 \left(-a^s \eta_5\right)^{\eta_4} \left(s \left(a^s \eta_5\right)^{\eta_2} \nu -\right. \nonumber \\
&&\left.e^{a^s \eta_5} \left(2 \mu -\nu  \left(-s \eta_4+a^s s \eta_6+\alpha  (2-\beta +\nu )\right)\right) \right.\nonumber \\
&&\left.\Gamma\left[\eta_2,a^s \eta_5\right]\right) \left(1-\left(s \left(a^s \eta_5\right)^{\eta_2} \left(-2 \mu +\nu  \left(-3-s \left(\eta_2+2 \eta_4-a^s (\eta_5+2 \eta_6)\right)+2 \alpha  (1-\beta +\nu )\right)\right)+\right.\right. \nonumber\\
&&\left.\left.e^{a^s \eta_5} \left(a^{2 s}
s^2 \eta_6^2 \nu +(3+s \eta_4+\alpha  (\beta -\nu )) (2 \mu +(s \eta_4+\alpha  (-2+\beta -\nu )) \nu )-\right.\right.\right. \nonumber\\
&&\left.\left.\left.a^s s \eta_6 (2 \mu +(3+s+2 s \eta_4+2 \alpha  (-1+\beta -\nu )) \nu )\right)
 \Gamma\left[\eta_2,a^s \eta_5\right]\right)\right.\nonumber \\
&&\left./ \left(3
\left(s \left(a^s \eta_5\right)^{\eta_2} \nu -e^{a^s \eta_5} \left(2 \mu -\nu  \left(-s \eta_4+a^s s \eta_6+\alpha  (2-\beta +\nu )\right)\right) \Gamma\left[\eta_2,a^s \eta_5\right]\right)\right)\right).
\end{eqnarray}

\begin{figure}[ht] \begin{minipage}[b]{0.45\linewidth} \centering\includegraphics[width=\textwidth]{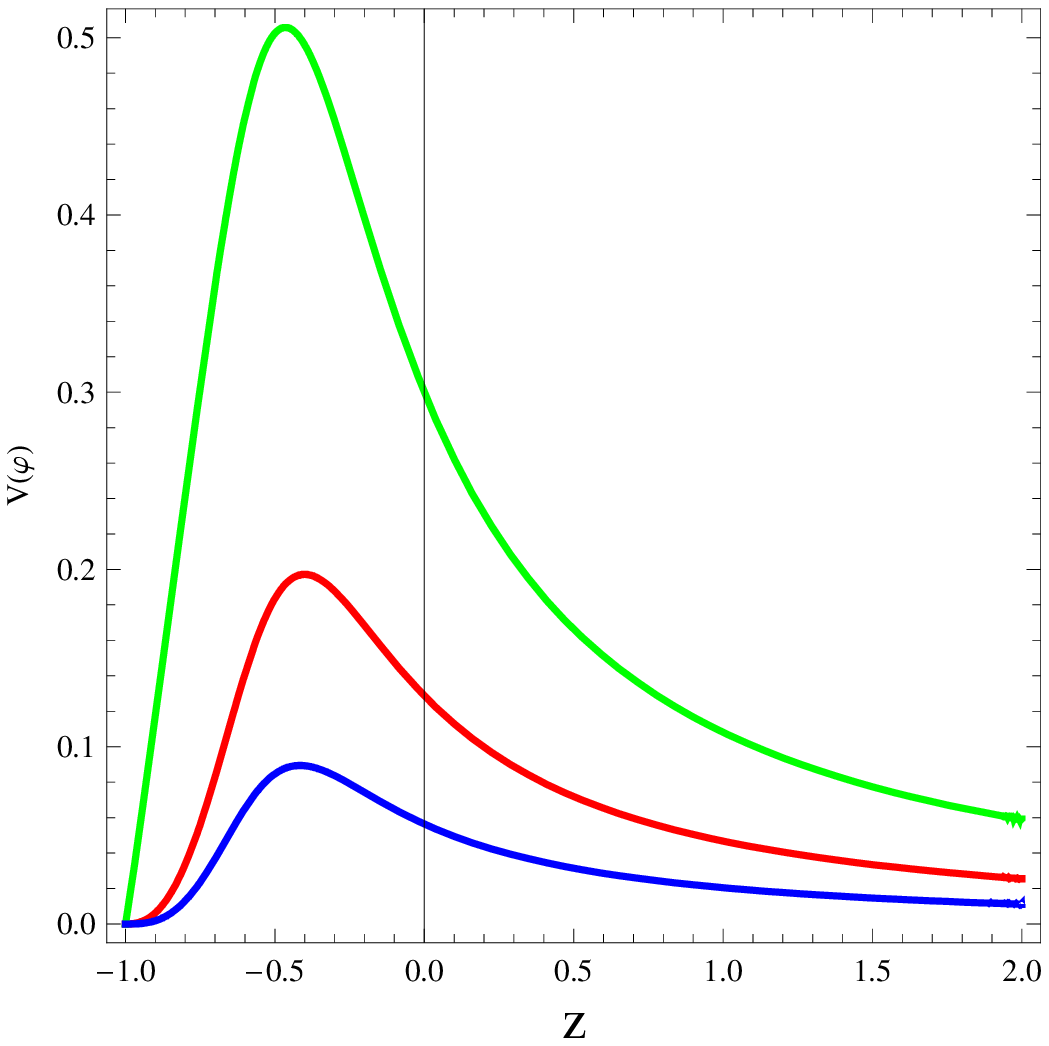} \caption{Plot of the evolution of the reconstructed potential $V(\varphi)$ for the reconstructed quintessence dark energy model. See Eq. (\ref{qpotential}).} \label{fig6} \end{minipage} \hspace{0.5cm} \begin{minipage}[b]{0.45\linewidth} \centering\includegraphics[width=\textwidth]{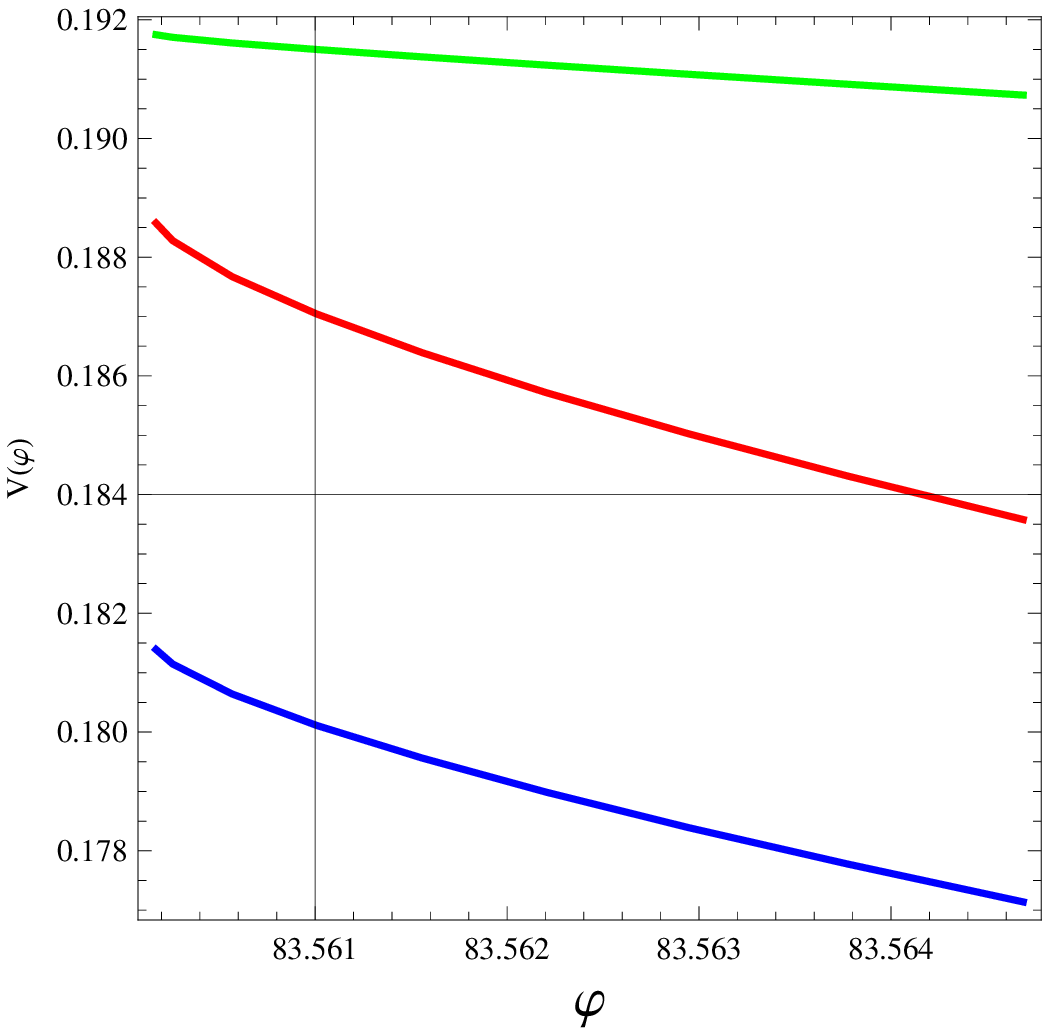} \caption{Plot of the evolution of the reconstructed potential $V(\varphi)$ (Eq. (\ref{qpotential})) with reconstructed scalar field $\varphi$ (Eq. (\ref{qfield})) of the quintessence dark energy model.} \label{fig7} \end{minipage} \end{figure}

\subsection{Reconstruction of DBI-essence dark energy model}
During the last few years, there have been many works aiming at connecting string theory with inflation which is also a phase
of accelerated expansion. Martin and Yamaguchi \cite{martin} introduced a scalar-field model of dark energy with a non-standard Dirac-Born-Infeld (DBI) kinetic term. This model is dubbed as ``DBI-essence dark energy" and the energy density $\rho_{DBI}$ and the pressure $p_{DBI}$ of the DBI-essence model are given, respectively, by \cite{martin}:
\begin{eqnarray}\label{dbi}
\rho_{DBI}&=&(\eta-1)T(\varphi)+V(\varphi),~~~~~~~~~~~~~~~~~~\\
p_{DBI}&=&\left(\frac{\eta-1}{\eta}\right)T(\varphi)-V(\varphi),~~~~~~~~~~~~~~~~~~
\end{eqnarray}
where:
\begin{eqnarray}
\eta=\frac{1}{\sqrt{1-\frac{\dot{\varphi}^2}{T}}}.
\end{eqnarray}
The EoS parameter for the DBI-essence scalar field model can be written as follow:
\begin{equation}\label{dbieos}
w_{DBI}= \frac{p_{DBI}}{\rho_{DBI}}  = \frac{\left(\eta-1\right)T(\varphi)-V(\varphi)\eta}{\eta\left((\eta-1)T(\varphi)+V(\varphi)\right)}.
\end{equation}
In the present work, we shall assume
 that $T=n\dot{\varphi}^2$, where $n>0$.
  Since we are considering a correspondence between DBI-essence dark energy and the reconstructed NHDE model, we consider $\rho_{DBI}=\rho_D$ and $w_{DBI}=w_D$. The based on Eq. (\ref{dbi}) we get the reconstructed scalar field $\varphi$ as function of the scale factor $a$ as follow:
\begin{eqnarray}
\label{dbiphi}
\dot{\varphi}(a)^2&=&-\frac{3^{\eta_4} 8^{-1-\eta_4}a^{\alpha  (\beta -\nu )}}{\omega }\times \nonumber \\
 && e^{-a^s (\eta_5+\eta_6)} \sqrt{1-\frac{1}{n}}
\phi_0^2 \eta_1 \eta_3 \left(-a^s \eta_5\right)^{\eta_4} \left(s \left(a^s \eta_5\right)^{\eta_2} \left(-2 \mu +\nu  \left(-s \left(\eta_2+2 \eta_4-a^s (\eta_5+2 \eta_6)\right)+\right.\right.\right. \nonumber \\
&&\left.\left.\left.\alpha  (2-2 \beta +2 \nu
)\right)\right)+e^{a^s \eta_5} \left(\alpha ^2 \nu  \left(\beta ^2-2 \beta  (1+\nu )+\nu  (2+\nu )\right)+s \left(s \eta_4^2 \nu
+2 \eta_4 \left(\mu -a^s s \eta_6 \nu \right)+\right.\right.\right.\nonumber \\
&&\left.\left.\left.a^s \eta_6 \left(-2 \mu +s \left(-1+a^s \eta_6\right) \nu \right)\right)+\right.\right. \nonumber\\
&&\left.\left.2 \alpha  \left(\beta  \left(\mu +s \left(\eta_4-a^s \eta_6\right) \nu \right)-
\nu  \left(\mu +s \left(\eta_4-a^s \eta_6\right) (1+\nu )\right)\right)\right) \Gamma\left[\eta_2,a^s \eta_5\right]\right),
\end{eqnarray}

\begin{eqnarray}
\label{dbiV}
V(\varphi(a))&=&
\frac{3^{\eta_4} 8^{-1-\eta_4}}{\omega } a^{\alpha  (\beta -\nu )} e^{-a^s (\eta_5+\eta_6)} \phi_0^2
\eta_1 \eta_3 \left(-a^s \eta_5\right)^{\eta_4}\times \nonumber\\
 &&\left(-3 \left(s \left(a^s \eta_5\right)^{\eta_2} \nu -e^{a^s \eta_5} \left(2 \mu -\nu  \left(-s \eta_4+a^s s \eta_6+\alpha  (2-\beta +\nu )\right)\right) \Gamma\left[\eta_2,a^s \eta_5\right]\right)+\right.\nonumber \\
&&\left.\left(-1+\sqrt{\frac{n}{n-1}}\right) \sqrt{\frac{n-1}{n}} n \left(s \left(a^s \eta_5\right)^{\eta_2} \left(-2 \mu +\nu  \left(-s \left(\eta_2+2 \eta_4-a^s (\eta_5+2 \eta_6)\right)+\right.\right.\right.\right. \nonumber\\
&&\left.\left.\left.\left.\alpha  (2-2 \beta +2 \nu
)\right)\right)+e^{a^s \eta_5} \left(\alpha ^2 \nu  \left(\beta ^2-2 \beta  (1+\nu )+\nu  (2+\nu )\right)+s \left(s \eta_4^2 \nu
+2 \eta_4 \left(\mu -a^s s \eta_6 \nu \right)+\right.\right.\right.\right. \nonumber \\
&&\left.\left.\left.\left.a^s \eta_6 \left(-2 \mu +s \left(-1+a^s \eta_6\right) \nu \right)\right)+2
\alpha  \left(\beta  \left(\mu +s \left(\eta_4-a^s \eta_6\right) \nu \right) \right. \right. \right. \right. \nonumber \\
&& \left. \left. \left. \left. -\nu  \left(\mu +s \left(\eta_4-a^s \eta_6\right) (1+\nu )\right)\right)\right) \Gamma\left[\eta_2,a^s \eta_5\right]\right)\right).
\end{eqnarray}

\begin{figure}[ht] \begin{minipage}[b]{0.45\linewidth} \centering\includegraphics[width=\textwidth]{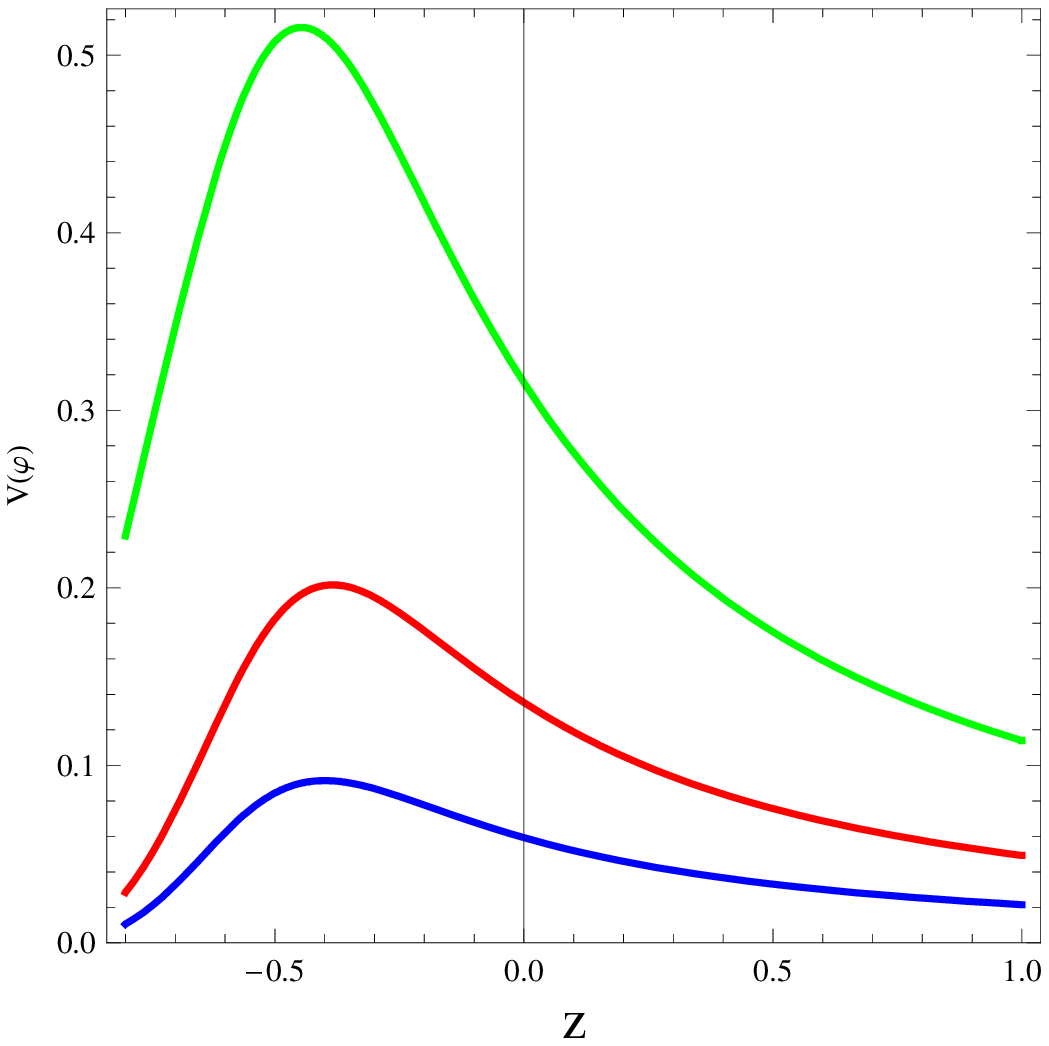} \caption{Plot of the evolution of the reconstructed potential $V(\varphi)$ for the reconstructed DBI-essence dark energy model. See Eq. (\ref{dbiV}). We have chosen $n=1.5$.} \label{fig9} \end{minipage} \hspace{0.5cm} \begin{minipage}[b]{0.45\linewidth} \centering\includegraphics[width=\textwidth]{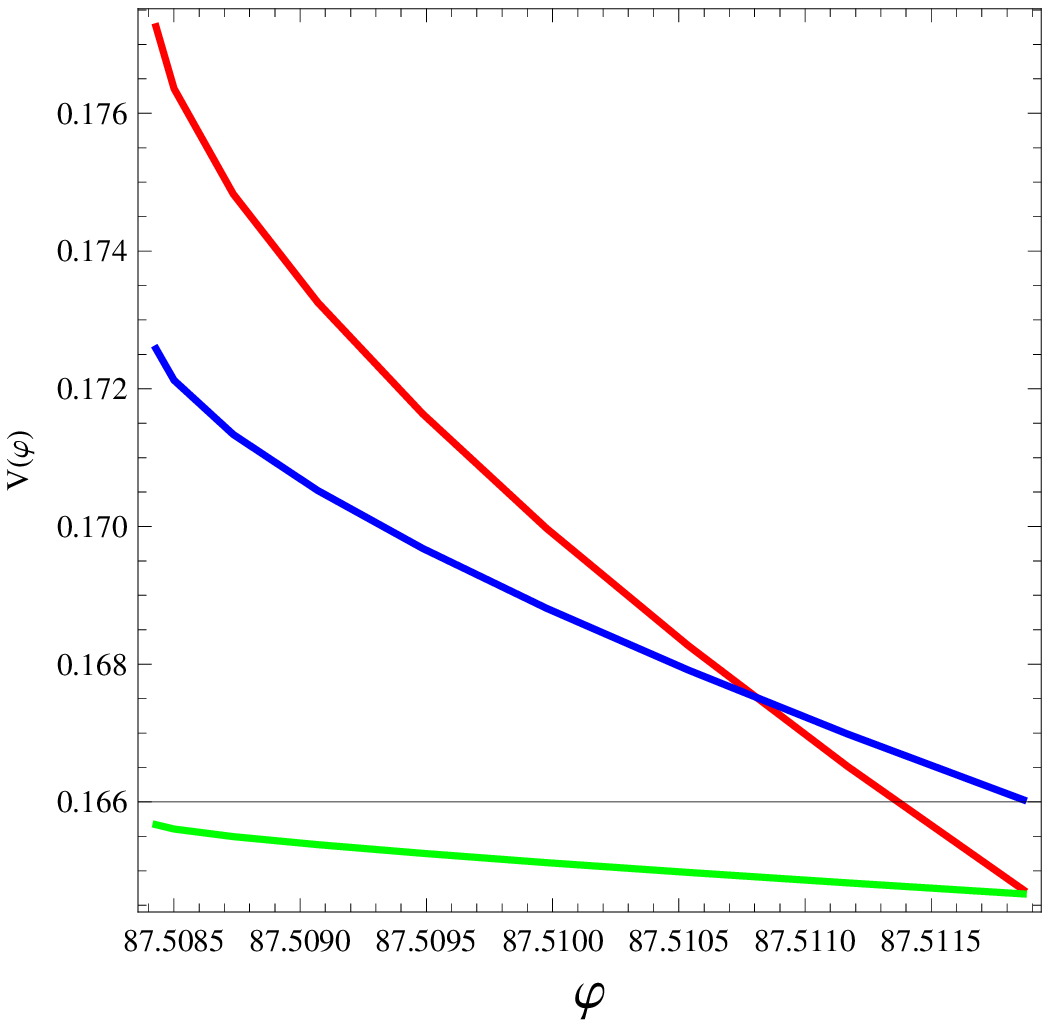} \caption{Plot of the evolution of the reconstructed potential $V(\varphi)$ (Eq. (\ref{dbiV})) with reconstructed scalar field $\varphi$ (Eq. (\ref{dbiphi})) of the DBI-essence dark energy model.We have chosen $n=1.5$.} \label{fig10} \end{minipage} \end{figure}

\subsection{Reconstruction of tachyon dark energy model}
Tachyonic condensate in a class of string theories can be described
by an effective scalar field with a Lagrangian of the form
$L=-V(\phi)(1-\partial_a\phi \partial^a \phi)^{1/2}$. Since this Lagrangian has also a potential
function $V(\phi)$, any form of cosmological evolution (that is, any $a(t)$) can be obtained with the tachyonic field as the source by
choosing $V(\phi)$ suitably \cite{paddy1}. Cosmological effects of homogeneous tachyon matter coexisting with
non-relativistic matter and radiation have been studied by \cite{paddy2}. The energy density $\rho_T$ and the pressure $p_T$ of the tachyon scalar field model are given, respectively, by \cite{paddy1}:
\begin{eqnarray}\label{tachyon}
\rho_T &=& \frac{V(\varphi)}{\sqrt{1-\dot{\varphi}^2}},\\
p_T &=& -V(\varphi)\sqrt{1-\dot{\varphi}^2},
\end{eqnarray}
while the  EoS parameter can be written as follow:
\begin{equation}\label{tacheos}
w_T=  \frac{p_T}{\rho_T}=  \dot{\varphi}^2-1.
\end{equation}
For the correspondence under consideration, we have $\rho_T=\rho_D$ and $w_D=w_T$. Using the same procedure used before, we reconstruct the scalar field $\varphi$ and the potential $V(\varphi)$ as follow:
\begin{eqnarray}
\label{tachphi}
\dot{\varphi}(a)^2&=&1+\left(s \left(a^s \eta_5\right)^{\eta_2} \left(-2 \mu +\nu  \left(-3-s \left(\eta_2+2 \eta_4-a^s
(\eta_5+2 \eta_6)\right)+\right.\right.\right. \nonumber\\
&&\left.\left.\left.2 \alpha  (1-\beta +\nu )\right)\right)+e^{a^s \eta_5} \left(a^{2 s} s^2 \eta_6^2 \nu
+(3+s \eta_4+\alpha  (\beta -\nu ))\times\right.\right. \nonumber\\
&&\left.\left.(2 \mu +(s \eta_4+\alpha  (-2+\beta -\nu )) \nu )-a^s s \eta_6 (2 \mu +(3+s+2 s \eta_4+2 \alpha  (-1+\beta -\nu )) \nu )\right) \Gamma\left[\eta_2,a^s \eta_5\right]\right)\times \nonumber\\
&&\left(3 \left(s \left(a^s \eta_5\right)^{\eta_2} \nu -e^{a^s \eta_5} \left(2 \mu -\nu  \left(-s \eta_4+a^s s \eta_6+\alpha  (2-\beta +\nu
)\right)\right) \Gamma\left[\eta_2,a^s \eta_5\right]\right)\right)^{-1},
\end{eqnarray}

\begin{eqnarray}
\label{tachV}
V(\varphi(a))&=&\left[-\frac{ 3^{1+2 \eta_4} a^{2 \alpha  (\beta -\nu )} e^{-2 a^s (\eta_5+\eta_6)} \phi_0^4 \eta_1^2 \eta_3^2}{\omega ^2}\times2^{-6 (1+\eta_4)}\right. \nonumber\\
&&\left. \left(-a^s \eta_5\right)^{2 \eta_4} \left(s \left(a^s \eta_5\right)^{\eta_2} \left(-2 \mu +\nu  \left(-3-s \left(\eta_2+2 \eta_4-a^s (\eta_5+2 \eta_6)\right)+2
\alpha  (1-\beta +\nu )\right)\right)+\right.\right.\nonumber \\
&&\left.\left.e^{a^s \eta_5} \left(a^{2 s} s^2 \eta_6^2 \nu +(3+s \eta_4+\alpha  (\beta -\nu ))
(2 \mu +(s \eta_4+\alpha  (-2+\beta -\nu )) \nu )-\right.\right.\right. \nonumber\\
&&\left.\left.\left. a^s s \eta_6 (2 \mu +(3+s+2 s \eta_4+2 \alpha  (-1+\beta -\nu )) \nu
)\right) \Gamma \times \right.\right. \nonumber \\
&&\left.\left. \left[\eta_2,a^s \eta_5\right]\right) \left(s \left(a^s \eta_5\right)^{\eta_2} \nu -e^{a^s
\eta_5} \left(2 \mu -\nu  \left(-s \eta_4+a^s s \eta_6+\right.\right.\right.\right.\nonumber\\
&&\left.\left.\left.\left.\alpha  (2-\beta +\nu )\right)\right) \Gamma\left[\eta_2,a^s \eta_5\right]\right)\right]^{1/2}
\end{eqnarray}

\begin{figure}[ht] \begin{minipage}[b]{0.45\linewidth} \centering\includegraphics[width=\textwidth]{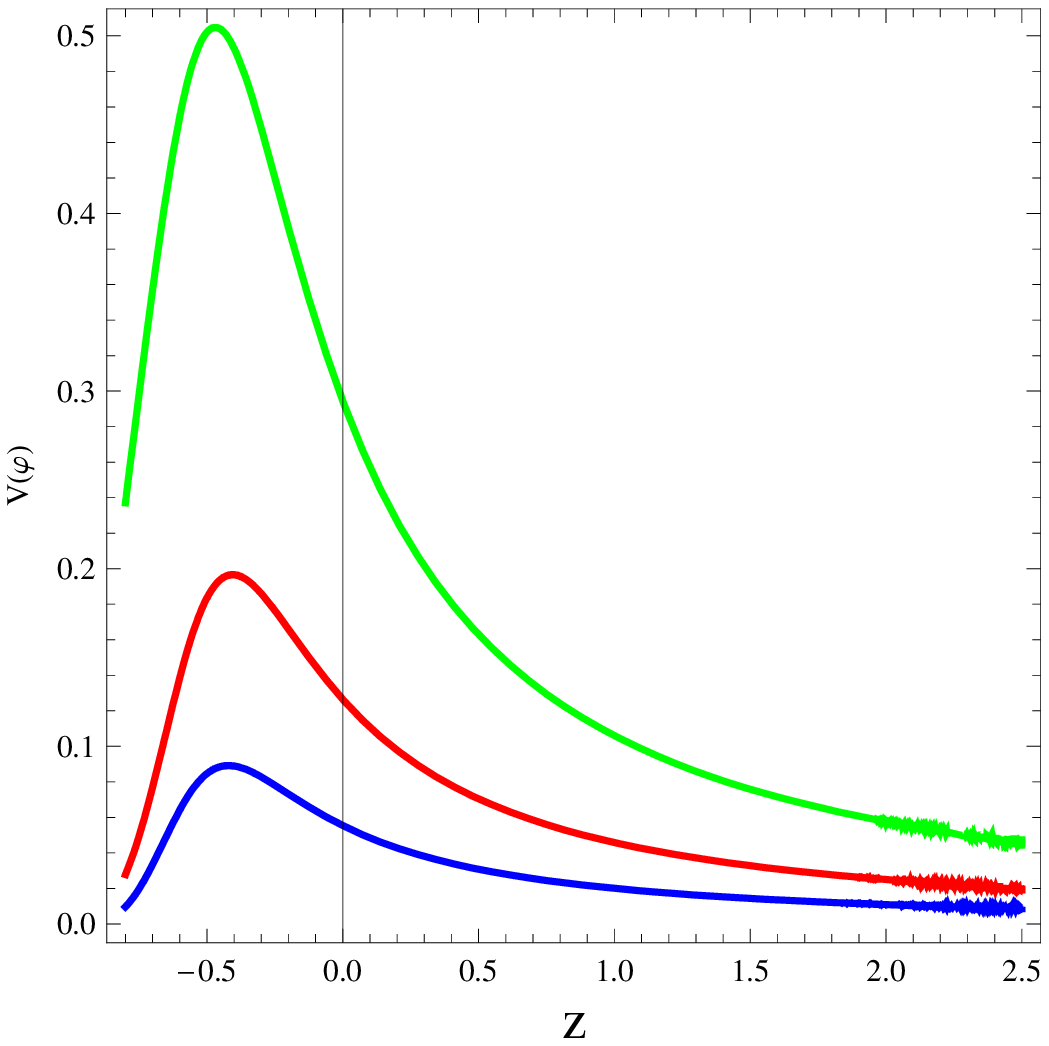} \caption{Plot of the evolution of the reconstructed potential $V(\varphi)$ for the reconstructed tachyon dark energy model. See Eq. (\ref{tachV}).} \label{fig11} \end{minipage} \hspace{0.5cm} \begin{minipage}[b]{0.45\linewidth} \centering\includegraphics[width=\textwidth]{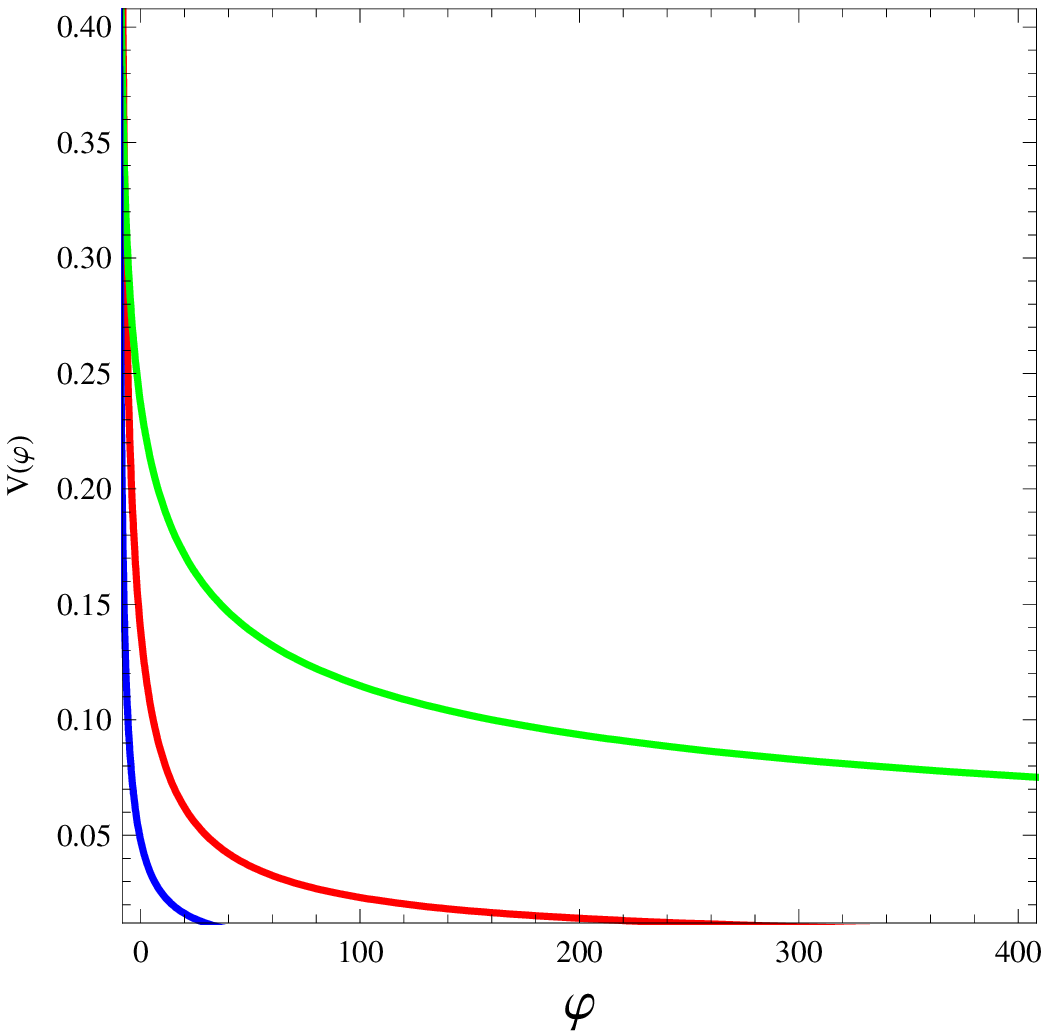} \caption{Plot of the evolution of the reconstructed potential $V(\varphi)$ (Eq. (\ref{tachV})) with reconstructed scalar field $\varphi$ (Eq. (\ref{tachphi})) of the tachyon dark energy model.} \label{fig12} \end{minipage} \end{figure}

\subsection{Discussion}
In this paper, we studied the main cosmological properties of the New Holographic Dark Energy (HNDE) model in the framework of Brans-Dicke chameleon cosmology. We considered a particular ansatz for the parameters $\phi$, $V$ and $f$ in which their expressions are given in the power law form.
We decided to consider different aspects to study. First of all, we reconstructed the expression of the Hubble parameter $H$ and, accordingly, the expression of the density $\rho_D$ of the NHDE  in the context of chameleon Brans-Dicke chameleon cosmology. We also tested the Weak Energy condition (WEC) and the Strong Energy Condition (SEC) for the reconstructed model we obtained. Considering three cases, namely $\{\mu=0.65,~\nu=0.20\}$, $\{\mu=0.60,~\nu=0.25\}$ and $\{\mu=0.55,\nu=0.15\}$ setting the other parameters as $\alpha=5,~\beta=-0.7,~\gamma=-0.9,~\phi_0=0.12,~\omega=-\frac{3}{2}+10^{-22},~f_0=1,~V_0=2$ and BD parameter $\omega$ following \cite{Hrycyna} we have computed the reconstructed EoS parameter. Observational results coming from SNeIa data suggest a limit of the EoS parameter as $-1.67<w<-0.62$ \cite{MNR}. Using a set values of $\mu$ and $\nu$ in Eq. (\ref{EoS}) we found that the results are in good agreement with observations of \cite{MNR}. Finally, we reconstructed three scalar field models of dark energy  (namely, the quintessence, the DBI-essence and the tachyon ones) based on the NHDE model in the framework of BD cosmology.  For the three scalar field models we considered, we have reconstructed the corresponding potentials and scalar fields. To further elucidate our reconstructions, we have plotted the reconstructed potential $V(\varphi)$ against $z$ and made parametric plots between $\varphi$ and $V({\varphi})$ in Figs. \ref{fig6}, \ref{fig7}, \ref{fig9}, \ref{fig10}, \ref{fig11} and \ref{fig12}. It is apparent from the plots that the potential $V(\varphi)$ is increasing up to redshifts of the order of  $z\approx -0.5$, afterwards it starts to decay. In the plots of $\varphi - V(\varphi)$, it appears that the potential has a decreasing behavior with the scalar field $\varphi$.

\begin{figure}[ht] \begin{minipage}[b]{0.45\linewidth} \centering\includegraphics[width=\textwidth]{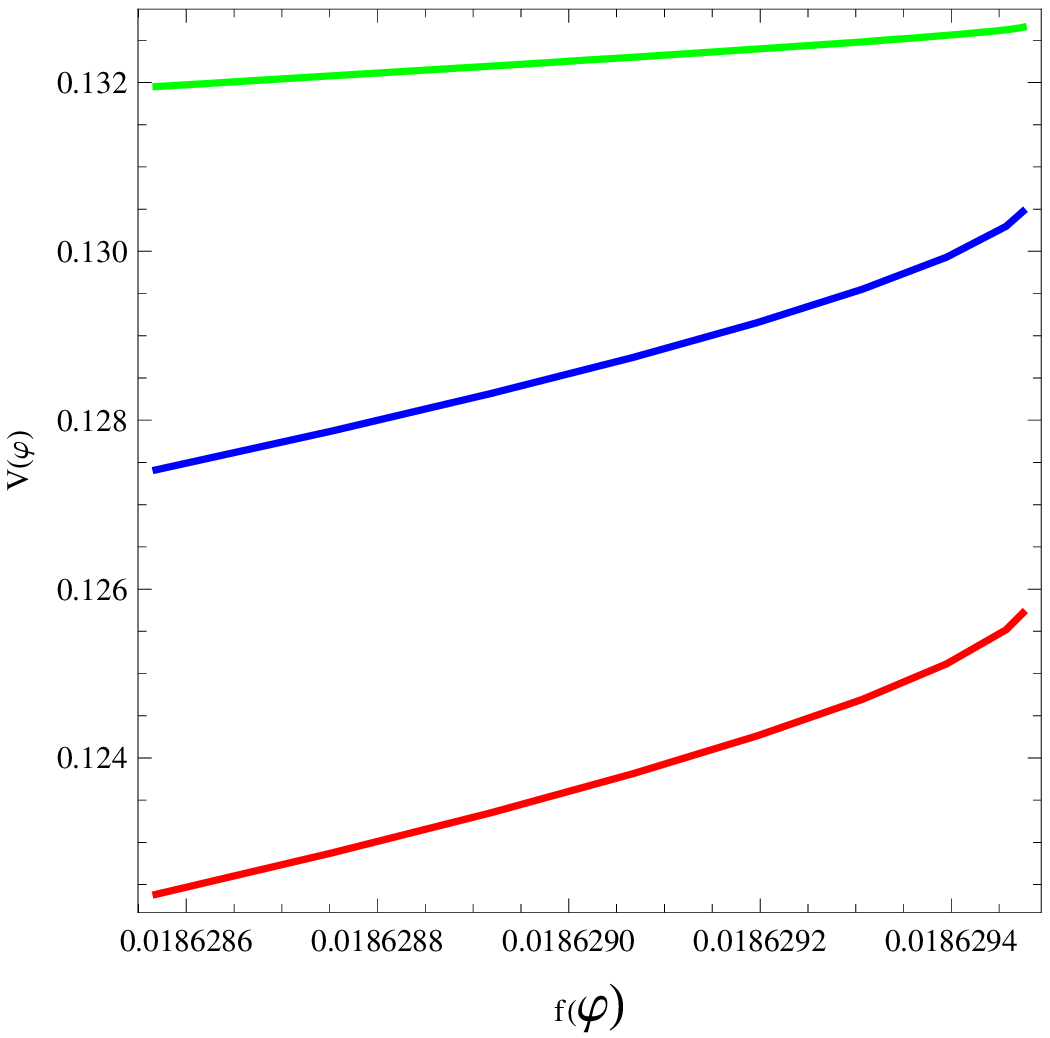} \caption{Plot of the evolution of the reconstructed potential $V(\varphi)$ with $f(\varphi)$ for quintessence dark energy model.} \label{quint} \end{minipage} \hspace{0.5cm} \begin{minipage}[b]{0.45\linewidth} \centering\includegraphics[width=\textwidth]{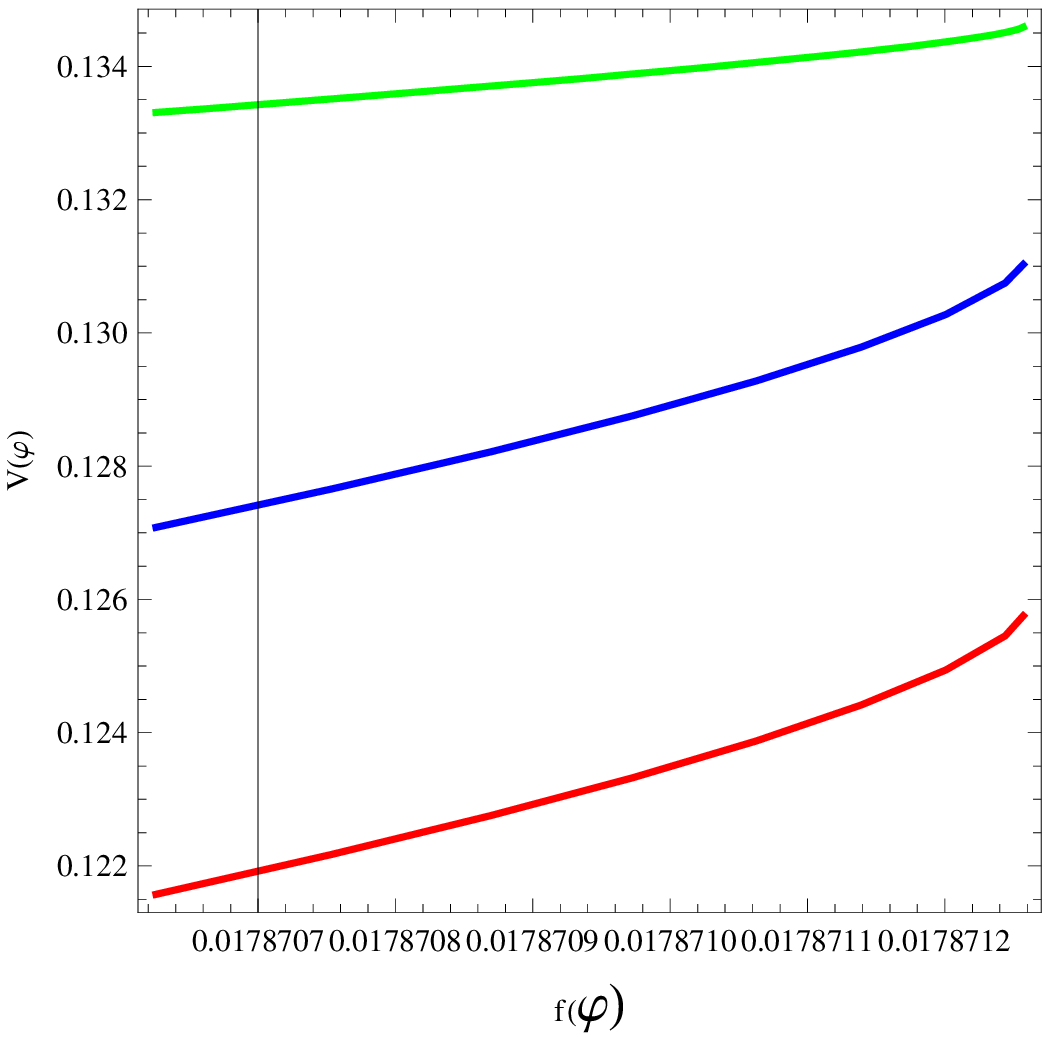} \caption{Plot of the evolution of the reconstructed potential $V(\varphi)$ with $f(\varphi)$ for DBI-essence dark energy model.} \label{dbi} \end{minipage} \hspace{0.5cm} \begin{minipage}[b]{0.45\linewidth} \centering\includegraphics[width=\textwidth]{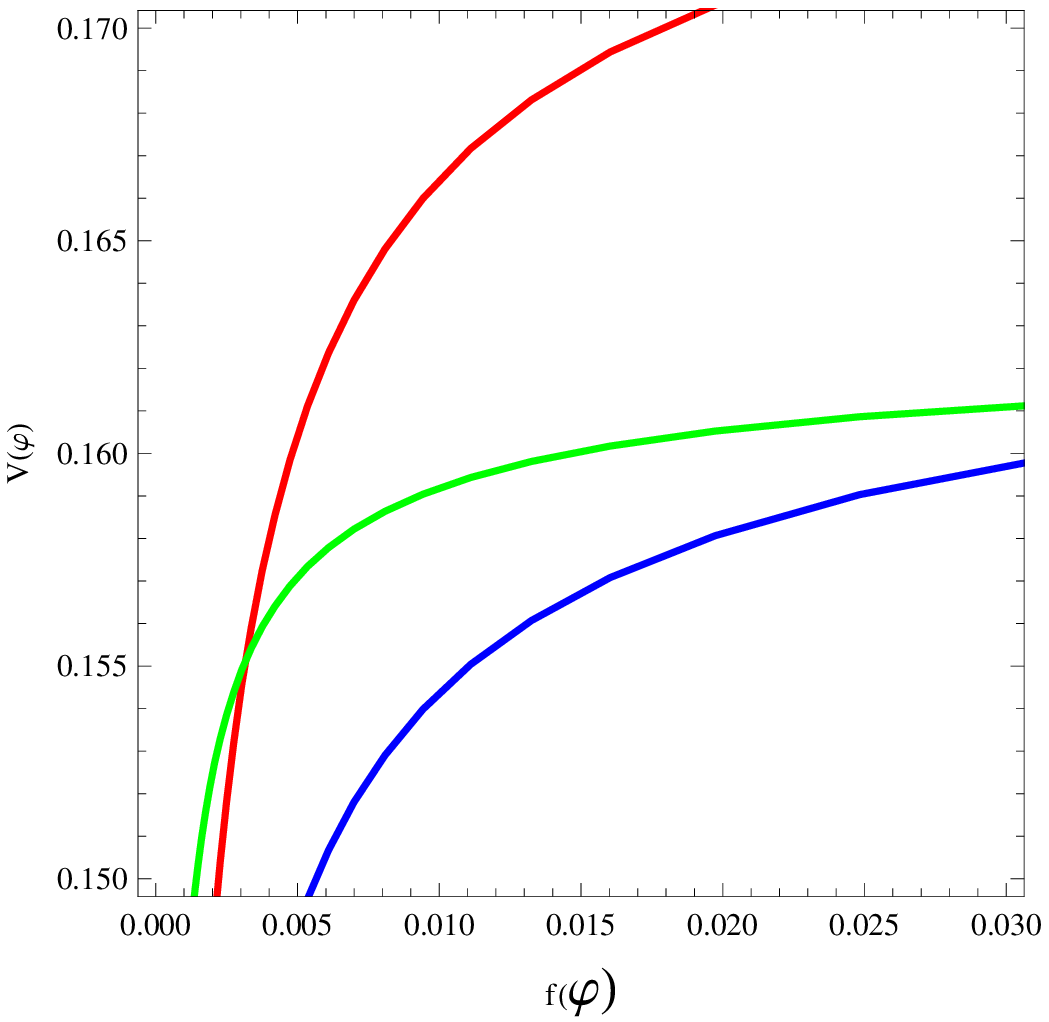} \caption{Plot of the evolution of the reconstructed potential $V(\varphi)$ with $f(\varphi)$ for tachyon dark energy model.} \label{tachyon} \end{minipage} \end{figure}

In order to have a look into the behavior of the reconstructed potential against coupling function $f$ we discuss the Figs. \ref{quint}, \ref{dbi} and \ref{tachyon}, where we observe that the reconstructed potentials are increasing with $f$. This indicates that the potential increasing as the matter-chameleon coupling is getting stronger. It is further noted that the rate of increase in the potential is much higher in the case of tachyon than the cases of quintessence and DBI.

\section{Concluding remarks}
In the present work we have used a reconstruction scheme for new holographic dark energy model with energy density given by $\rho_D=\frac{3 \phi^2 }{4 \omega }\left(\mu H^2  +\nu \dot{H}  \right)$ in the framework of Brans-Dicke cosmology taking the ansatz $\phi=\phi_0 a^\alpha,~~V=V_0 \phi^\beta,~~f=f_0 \phi^\gamma$. Outcomes of the study are:
\begin{itemize}
  \item Considering $\rho=\rho_D$ in the first modified field equation of BD theory leads to a linear differential equation which could be solved analytically to have a solution for the reconstructed Hubble parameter in terms of scale factor $a$; when plotted against redshift $z=a^{-1}-1$, it exhibited decaying pattern with the evolution of the universe (i.e., decreases in $z$) and this is consistent with the accelerated expansion of the universe.
  \item The NHDE energy density, as reconstructed through Hubble parameter, when plotted against $z$, is found to increase with evolution of the universe and it is consistent with the evolution of the universe from matter to dark energy domination.
  \item Violation of strong energy condition, as expected in the framework of Einstein gravity, has also been found for the reconstructed NHDE model in the framework of BD gravity.
  \item The reconstructed equation of state (EoS) parameter has been found to exhibit ``phantom"-like behavior, i.e. $w_D<-1$.
  \item Considering three different combinations of the parameters $\mu$ and $\nu$, namely $\{\mu=0.65,~\nu=0.20\}$, $\{\mu=0.60,~\nu=0.25\}$ and $\{\mu=0.55,\nu=0.15\}$ and setting the other parameters as $\alpha=5,~\beta=-0.7,~\gamma=-0.9,~\phi_0=0.12,~\omega=-\frac{3}{2}+10^{-22},~f_0=1,~V_0=2$ and the BD parameter $\omega$ following \cite{Hrycyna}, we have computed the reconstructed EoS parameter for the reconstructed NHDE. Observational results coming from SNeIa data suggest a limit of the EoS parameter as $-1.67<w<-0.62$ \cite{MNR}. Using a set of variations in the values of $\mu$ and $\nu$ in Eq. (\ref{EoS}) we found that the results are in good agreement with observations of \cite{MNR} (see Table I).
\end{itemize}
In the following phase of the study, we considered the correspondence between the reconstructed new holographic dark energy in the framework of BD gravity and some scalar field dark
energy models in a manner under which the two scenarios can
be simultaneously valid. This type of approach is available in cosmological literature (e.g., \cite{motiv1,motiv2,motiv3}). We have constructed the potentials and the scalar fields of these models. We observed that $\frac{dV}{dz}>0$ for all of the reconstructed scalar field models upt o $z\approx -0.5$ and, at very late stage, (i.e., $z<-0.5$), we have $\frac{dV}{dz}<0$. Moreover, $\frac{dV}{d\phi}<0$ for all of the models.

In summary, by generalizing the previous works \cite{motiv1,motiv2,motiv3} to the NHDE model with $\rho_D=\frac{3\phi^2}{4\omega}(\mu H^2+\nu \dot{H})$ in the framework of chameleon Brans-Dicke cosmology, we have obtained the evolution of EoS. Following \cite{motiv1,bisbar} we have considered $V$, $\phi$ and $f$ in power-law form and accordingly reconstructed Hubble parameter. This approach differs from \cite{motiv1} in the sense that instead of considering Brans-Dicke cosmology, we have considered chameleon Brans-Dicke with coupling function $f$. We have tested SEC and WEC conditions and interpreted evolution of EoS from them. With some choice of the model parameters we computed EoS and found the the computed values of EoS are consistent with the observational results coming from SNeIa data that suggest a limit of the EoS parameter as $-1.67<w<-0.62$ \cite{MNR}. Subsequently we examined the stability for the obtained solutions of the crossing of the phantom divide under a quantum correction of massless conformally-invariant fields and we have seen that quantum correction could be small when the phantom crossing occurs and the obtained solutions of the phantom crossing could be stable under the quantum correction. In the subsequent phase, we have established a correspondence between the NHDE model and the quintessence, the DBI-essence and the tachyon scalar field models in the framework of chameleon Brans-Dicke cosmology. We reconstruct the potentials and the dynamics for these three scalar field models we have considered. The reconstructed potentials are found to increase with evolution of the universe and in a very late stage they are observed to decay. It is also observed through $f(\varphi)-V(\varphi)$ plot that the potential is increasing with $f$, which indicates that the potential increases as the matter-chameleon coupling gets stronger with evolution of the universe.

\section{Acknowledgements}
Sincere thanks are due to the anonymous reviewer for constructive suggestions. Project Grant of DST, Govt. of India no. SR/FTP/PS-167/2011 is duly acknowledged by the first author. Also, the first authors acknowledges the facilities provided by IUCAA, Pune, India, where a major portion of the work was carried out during a scientific visit in December, 2013-January, 2014.

\end{document}